\documentclass{ws-mpla}

\usepackage{graphicx}
\usepackage{amssymb}

\begin{document}

\markboth{T. Padmanabhan}
{Topological interpretation of the horizon temperature}

\catchline{}{}{}{}{}
%
%%%%%%%%%%%%%%%%%%%%%%%%%%%%%%%%%%%%%%%%%%%%%%%%%%%%%%%%%%%%%%%%%%%%

\title{Topological interpretation of the horizon temperature
}

\author{\footnotesize T. Padmanabhan}

\address{Inter-University Centre for Astronomy and Astrophysics,\\
Post Bag 4, Ganeshkhind, Pune - 411 007, INDIA.\\
E-mail address: nabhan@iucaa.ernet.in}

\maketitle

\pub{Received (Day Month Year)}{Revised (Day Month Year)}

\begin{abstract}

\noindent A class of metrics $g_{ab}(x^i)$ describing  spacetimes with  horizons (and associated thermodynamics) can be thought of as a limiting case of a family of metrics $g_{ab}(x^i;\lambda)$ {\it without horizons} when  $\lambda\to 0$. I construct specific examples in which the curvature corresponding 
$g_{ab}(x^i;\lambda)$ becomes a  Dirac delta function and gets concentrated on the horizon
when the limit $\lambda\to 0$ is taken, but the action remains finite.  When the horizon is interpreted in this manner, one needs to remove the corresponding surface from the Euclidean sector, leading to winding numbers and thermal behaviour. 
In particular, the Rindler spacetime can be thought of as the limiting case of (horizon-free) metrics of the form
[$g_{00}=\epsilon^2+a^2x^2;g_{\mu\nu}=-\delta_{\mu\nu}$] or [$g_{00} = - g^{xx} = (\epsilon^2 +4 a^2 x^2)^{1/2}, 
g_{yy}=g_{zz}=-1]$ when $\epsilon\to 0$. In the Euclidean sector, the curvature gets concentrated on the origin of $t_E-x$ plane in a manner analogous to Aharanov-Bohm effect (in which the the vector potential is a pure gauge everywhere except at the origin) and the curvature  at the origin leads to nontrivial topological features and winding number.

\keywords{Horizon temperature; Aharonov-Bohm effect; Quantum gravity}

\end{abstract}

\ccode{PACS Nos.: 04.20.Gz, 04.60.-m, 04.62.+v.}

\section {Parametrisation dependence in Euclidean extension}

   In the case of the  Rindler, Schwarzschild and de Sitter  
    metrics (for which a thermodynamical interpretation can be provided),
      one can introduce\cite{birreltp} a set of coordinates
    $\mathcal{G} =\{ X^a \} $ which covers the entire spacetime manifold and another set of coordinates
    $\mathcal{S} = \{ x^a \}$ which covers only part of the manifold. 
    The metric is static with respect to $x^0$ since it is 
    associated with a Killing symmetry of the spacetime.
      Under the Euclidean extension of  $\mathcal{S}$, the resulting metric should exhibit periodicity
    in terms of the Euclidean time coordinate $\tau = i x^0$ leading to a non-zero temperature.
     
 For example, let us consider the case of flat spacetime represented in Minkowski and Rindler coordinates.
 Since the transformation between $\mathcal{G}$ and $\mathcal{S}$ in the case of many other spacetimes (like
 Schwarzschild, de Sitter etc) has identical  structure
    when the spacetime is embedded in higher dimension\cite{tprealms}, the results obtained in the 
    case of Minkowski and Rindler frames can be easily translated to other cases.
In the case of  
    the inertial ($T,{\bf X}$) and Rindler ($t,{\bf x}$) coordinates,
    related by the transformation $(-\infty< (T,X,t,x)<\infty)$
    \begin{equation}
    T=x \sinh at,  X= x \cosh at
    \label{rindlering}
    \end{equation}
     in the region $|X| > |T|$, the line element is
    \begin{equation}
    ds^2=dT^2 - d{\bf X}^2 = a^2x^2 dt^2 - d{\bf x}^2
    \end{equation}
    The Euclidean version of the transformation between ($T_E = i T,{\bf X}$) and 
    ($t_E=it,{\bf x}$) is given by $T_E=x \sin at_E, X= x \cos at_E$ which is 
    the standard transformation connecting Cartesian and Polar coordinates. The
    Euclidean metric,
    \begin{equation}
    -ds^2=dT_E^2 + d{\bf X}^2 = x^2 d(at_E)^2 + d{\bf x}^2
\label{basiceuclid}
    \end{equation}
    however, will have a conical singularity unless the coordinate $t_E$ has a periodicity
    $(2\pi/a)$; this  leads to the thermal effects.

     The  transformations (\ref{rindlering})
    between the Euclidean versions of $\mathcal{G}$ and $\mathcal{S}$  has some curious features which needs to be stressed:  

   (i) Consider a parametrised curve ($T(s),{\bf X}(s)$) in the inertial frame and its mapping under Euclidean
extension. In general, such a curve will not map to real values under the transformation ($T_E = i T,{\bf X}$). 
(A simple example is the straight line $X=vT,Y=Z=0$, which goes over to $X=-ivT_E,Y=Z=0$; this cannot be expressed in a {\it real} $T_E-X$ plane.)
Even when it is possible to represent a
    curve in both Minkowski and Euclidean sectors, its topological nature as well as range can vary significantly. For example, 
a {\it single} hyperbola of the form 
\begin{equation}
X=+\sqrt{1+ T^2}, \qquad (-\infty<T<+\infty) 
\end{equation}
in the Minkowski space 
     will go over to 
\begin{equation}
X=+\sqrt{1- T_E^2}, 
\label{hypereuclid}
\end{equation}
on analytic continuation. The original range of $(-\infty<T<+\infty)$ now gets mapped into
$(-\infty<T_E<+\infty)$.; but equation (\ref{hypereuclid}) shows that $X$ is real only for $|T_E|\le1$. Thus
the entire hyperbola cannot be represented in the {\it real} $T_E-X$ system. For the range $|T_E|\le1$,
when $X$ is real, the curve is a
      {\it semi}-circle. [For $|T_E|\ge1$, equation  (\ref{hypereuclid}) represents a pair of hyperbolas in the $( Im X,T_E)$
     plane for  which cannot  be drawn in the $(Re X,T_E)$ plane.)  

\begin{figure}   
   \begin{center}
   \includegraphics[angle=270,scale=0.4]{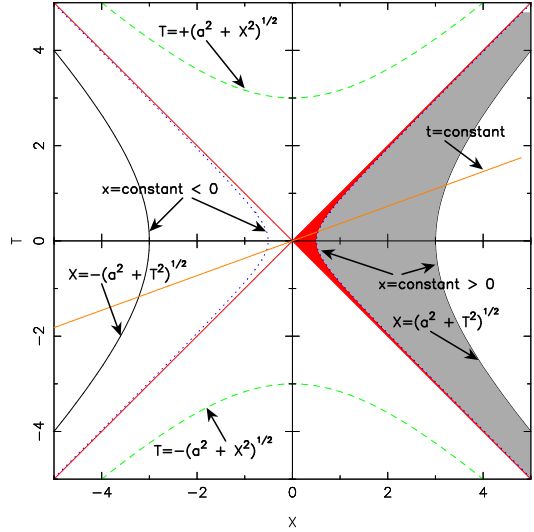}
   \end{center}
\caption{
The Rindler coordinate system ($t,x$) covers the right ($-\infty<t<\infty, x>0$)  and left ($-\infty<t<\infty, x<0$) wedges of the Minkowski manifold with ($-\infty<(X,T)<\infty$). The Euclidean section is shown in figure \ref{fig:rindlerfigsb}.
}
 \label{fig:rindlerfigsa}
\end{figure}

\begin{figure}   
  \begin{center}
  \includegraphics[angle=270,scale=0.4]{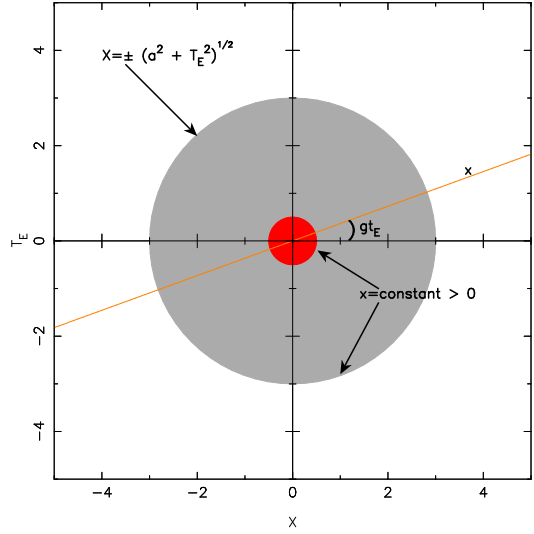}
   \end{center}
\caption{
The Euclidean extension of Minkowski manifold with ($-\infty<(X,T_E)<\infty$) is covered by the ``polar
coordinates" $(x,t_E)$ with $0<at_E<2\pi$. The shaded regions in figure \ref{fig:rindlerfigsa} maps to the
corresponding shaded regions in this figure. In particular, the range $\pi/2<at_E<3\pi/2$ with $x>0$ can
cover the $X<0$ wedge of the Minkowski coordinate.
}
 \label{fig:rindlerfigsb}
\end{figure}

(ii) Now consider the same,
     {\it single} hyperbola, given in parametric form as $T=\sinh t,  X= \cosh t$. On analytically continuing in
     both $T$ and $t$, we get  $T_E=\sin t_E, X= \cos t_E$ with $\infty<t_E<+\infty$. This shows that the same {\it single} hyperbola gets mapped to the {\it full} circle in the $(X,T_E)$ plane. [We stress the fact that nowhere did we invoke the hyperbola in the left wedge;
     see figures (\ref{fig:rindlerfigsa}) and (\ref{fig:rindlerfigsb})]. The  mapping $T_E=\sin t_E$ is many-to-one
and limits the range of $T_E$  to $|T_E|\le1$ for $(-\infty < t_E < \infty)$.
The key new feature is the analytic continuation in the parameter $t$ as well.

(iii) The above result translates to the following feature in the transformation between inertial and 
Rindler coordinates. The transformations in (\ref{rindlering}) with $x>0, -\infty<t<\infty$ covers \emph{only} the right
    hand wedge [$|X|>|T|, X>0$] of the Lorentzian sector; one needs to take $x<0, -\infty<t<\infty$ to cover the
    left hand wedge  [$|X|>|T|, X<0$]. The metric in 
(\ref{basiceuclid}), interpreted in analogy with polar coordinates, however,
   has
    $x>0$. {\it Nevertheless, both $X>0$ and $X<0$ are covered by different ranges of the ``angular" coordinate $t_E.$} The range $(-\pi/2) < at_E <(\pi/2)$ [right half of the circle in figure (\ref{fig:rindlerfigsb})] covers $X>0$ while the range $(\pi/2) < at_E <(3\pi/2)$
   (left half of the circle) covers $X<0$.  Thus the Euclidean sector ``knows'' about the region beyond the horizon
    (the left wedge) even though $x>0$. 

(iv) The left half of the circle can  arise purely from the part of the  hyperbola in the {\it right wedge} corresponding to $(\pi/2) < at <(3\pi/2)$ and it is {\it not} necessary to introduce the hyperbola in the left wedge to get the left half of the circle. More generally, all the  events $\mathcal{P}_n\equiv(t=(2\pi n/a),{\bf x})$  [where $n = 
    \pm 1,\pm 2, ...$] which
    correspond to {\it different} values of $T$ and $X$  will be mapped to the {\it same} point in the 
    Euclidean space. Thus the circle in figure  (\ref{fig:rindlerfigsb}) is traversed several times as one moves along the hyperbola in figure (\ref{fig:rindlerfigsa}). (iv) The light cones of the inertial frame $X^2=T^2$ are mapped into the horizon $x=0$ in the Rindler frame and to the origin of the figure (\ref{fig:rindlerfigsb}) in the $T_E-X$ plane. The region ``inside" the horizon
    $|T|>|X|$ simply \emph{disappears} in the Euclidean sector.  Region  within one Planck length, $L_P$
    of the horizon, say, will be confined to a circle of radius $L_P$ around the origin in the Euclidean plane.

%%%%%%%%%%%%%%%%%%%%        FIGURE  %%%%%%%%%%%%%%%%%%%%%%

%%%%%%%%%%%%%%%%%%%%%%%%%%%%%%%%%%%%%%%%%%%%%%%%%%    

%%%%%%%%%%%%%%%%%%%%% FIGURE %%%%%%%%%%%%%%%

%%%%%%%%%%%%%%%%%%%%%%%%%%%%%%%%%%%%%%%%

\section{ Mission Impossible: Changing topology by changing coordinates}

   Since the horizon $x=0$ [or $X^2=T^2$] plays an important role in the thermodynamics of the spacetime,
    it is natural to explore whether one can attribute any special properties to the origin in the $T_E-X$ plane.
    One could, for example, take the point of view that since the region bounded by the horizon is inaccessible to the Rindler observer, the origin should be
    removed from the Euclidean spacetime. This will change the topological structure and will introduce a nontrivial winding number for paths that go around the origin. The thermal effect can be obtained directly from this feature.\cite{chrisduff} The conceptual difficulty  with this approach is that, Euclidean Rindler coordinates are the same as the standard polar coordinates in a plane; removing the origin is then like removing the origin from a two dimensional, flat, sheet of paper just because one is using polar coordinates. That hardly seems justifiable.
At the same time, the economy and geometrical beauty of the above argument is so attractive that I shall provide an alternative way of generalizing and interpreting this result.

To do this, let us begin with a class of metrics of the form
\begin{equation}
ds^2 = f (x) dt^2 - dx^2 - dy^2-dz^2
\label{startmetric}
\end{equation}
where $f$ is an even function of $x$. [This restriction is not essential but makes the characterization of some of the results easy]. Using the scalar curvature for these metrics
\begin{equation}
R=-\frac{f^{'2} - 2 f f''}{2f^2} = \frac{1}{\sqrt{f}} \frac{d}{dx} \left(  \frac{1}{\sqrt{f}} \frac{df}{dx}\right)
\end{equation}
we can compute the Einstein Hilbert action to be
  \begin{equation}
A = -\frac{1}{16\pi} \int_0^\beta dt \int_{S_\perp} d^2 x_\perp \int_{x_1}^{x_2} dx  
  \sqrt{f}  R 
  =-\frac{\beta S_\perp}{16\pi}\left[ \frac{f'}{\sqrt{f}} \right]_{x_1}^{x_2}
\label{actionval}
\end{equation}
Since $R$ is independent of $t$ and the transverse coordinates we have to restrict the integration
over these to a finite range with $0\le t \le \beta$ and $S_\perp$ being the transverse area.  The integral over $x$ is taken over the range $x_1 < x <x_2$. Let us now confine attention to a subset of $f$ which are {\it asymptotically Rindler}, in the sense that $f(x^2) \to a^2 x^2 $ as $ x^2 \to \infty)$
For these metrics, when $x_1\to-\infty,x_2\to+\infty$, the action is given by
\begin{equation}
A =-2\left( \frac{S_\perp}{4}\right) \left(\frac{a\beta}{2\pi}\right) 
\label{actionvalue}
\end{equation}
which is completely independent of the detailed behaviour of $f(x)$ at finite $x$.  
[Note that for $x\to \pm\infty, f'\to (2a^2x)$ but $ \sqrt{f}\, \to a|x|$; so, $(f'/\sqrt{f})$ in  (\ref{actionval})  
gives $2a$ at $x=+\infty$ but $-2a$ at $x=-\infty$.] 
This is a special case of a more general result. For any static spacetime with a metric
\begin{equation}
ds^2 = N^2({\bf x}) dt^2 - \gamma_{\alpha\beta}({\bf x})dx^\alpha dx^\beta
\end{equation}
 we have $R = {}^3R+2 \nabla_i a^i$ where $a_i=(0,\partial_\alpha N/N)$ is the acceleration of ${\bf x} = $ constant world lines. Then, limiting the time integration to $(0,\beta)$, say, the action becomes
\begin{equation}
A =- \frac{\beta}{16\pi}\int_{\mathcal{V}}d^3 x N\sqrt{\gamma}{}^3R-
 \frac{\beta}{8\pi} \int_{\partial\mathcal{V}} (a^\alpha n_\alpha)N\sqrt{\sigma} d^2x
\end{equation}
where $\sigma_{ab}$ is the induced metric on the two dimensional boundary $\partial\mathcal{V}$.
For metrics with ${}^3R\to 0$, the action  depends only on the surface gravity of the boundary $\partial \mathcal{V}$ of $\mathcal{V}$.

 Let us now write $f(x^2)=F(x^2)+a^2x^2$ with $F(x^2)\to 0$ as $x^2\to\infty$. Since the result in (\ref{actionvalue})
holds independent of $F$, it will continue to hold even when we take the limit of $F$ tending to zero. But when $F$
goes to zero, the metric reduces to standard Rindler metric and one would have expected the scalar curvature to vanish identically, making $A$ vanish identically. Our result in (\ref{actionvalue})    shows that the action is finite even for a Rindler spacetime  \emph{if we interpret it as arising from the limit of these class of metrics.}
It is obvious that, treated in this limiting fashion, as $F$ goes to zero $R$ should become a distribution in $x^2$ 
such that it zero almost everywhere except at the origin and has a finite integral. To see how this comes 
about,  let us study an explicit example. 

Consider the class of two parameter metrics with $f(x)=\epsilon^2+a^2x^2$. 
 When $a=0$ this metric represents flat spacetime in  standard Minkowski coordinates; it  also represents flat spacetime for $\epsilon=0$ but now in the Rindler coordinates. 
For finite values of $(\epsilon,a)$ the spacetime is curved with
\begin{equation}
R = \frac{2\epsilon^2 a^2}{(\epsilon^2 + a^2 x^2)^2} = - \frac{1}{2} R^t_{\phantom{t}x tx}
\label{reqn}
\end{equation}
[Only the component $R_{tx tx}$ and those related to it by symmetries of the curvature tensor
are non zero for this metric.]
There is no horizon when $\epsilon\ne 0$. When $\epsilon\ne 0, a\to 0$ limit is taken, we obtain the flat spacetime in Minkowski coordinates without ever producing a horizon. But the limit $a\ne 0,\epsilon\to0$ leads to a different result:
when $\epsilon\to 0$, a horizon appears at $x=0$. To study the properties, we use the limit 
\begin{equation}
\lim_{\epsilon\to 0}\frac{\epsilon^2}{(\epsilon^2+l^2)^2}=\delta(l^2)
\label{implimit}
\end{equation}
valid for $0\le l <\infty$. [A  rigorous proof of this result  can be found in Ref.~\refcite{stackgold}
and is based on the following theorem: If $F(r)$ is a unit normalized function, then the sequence of functions
$F_\mu(r)=\mu^{-1}F(r/\mu)$ tends to the Dirac delta function when $\mu\to0$.]
In this limit,
the scalar curvature $R$ 
in (\ref{reqn}) becomes   the distribution
\begin{equation}
\lim_{\epsilon\to 0} R = 2 \delta(x^2) 
\label{rlimiteqn}
\end{equation}
showing that the curvature is concentrated on the surface $x^2=0$ giving a finite value to the action even though
the metric is almost everywhere flat in the limit of $\epsilon\to 0$. In this limit we can introduce the Minkowski coordinates
both in spacetime and in its Euclidean extension. The entire analysis goes through even in the Euclidean sector, showing that the curvature is concentrated on  $x^2=X^2+T_E^2=0$. When analytically continued to the Lorentzian
sector, the curvature is on the light cones $x^2=X^2-T^2=0$. Thus, if we treat the Rindler frame as a limit of a sequence of metrics with $g_{00}=(\epsilon^2+a^2x^2)$, then it makes sense to exclude the origin from the Euclidean plane or the horizon from the spacetime. This will lead to a nontrivial topological winding number and 
a topological interpretation of the thermal behaviour.\cite{chrisduff}

Similar conclusions can be obtained from another class of metrics: Let
\begin{equation}
ds^2 =h(l)dt^2 - \frac{dl^2}{h(l)} - dy^2-dz^2
\label{classofmetrics}
\end{equation}
where 
\begin{equation}
h(l)=k_2+k_1l+\epsilon F(l/\epsilon)
\end{equation}
 with $k_1,k_2$ being constants $F$ being an arbitrary function of its argument, subject to the condition
that integral of $F''[z]$ with respect to $z$ over the real line is finite (and equal to $I$, say). In this case,
the scalar curvature and the action [in the notation of equation (\ref{actionvalue})]
are  given by 
\begin{equation}
R=d^2h/dl^2=\epsilon^{-1}F''(x/\epsilon);\qquad A=-(\beta S_\perp/16\pi)I
\end{equation}
Note that the action is   independent of $\epsilon$. In the limit of $\epsilon\to 0$, the curvature
goes to $R\propto\delta_D(x)$, while we can easily choose $F$ such that as $\epsilon\to0$, 
\begin{equation}
\epsilon F(x/\epsilon)\to (k_3x),\qquad h(l)\to [k_2+(k_1+k_3) x]
\end{equation}
 which is a Rindler coordinate system for $[k_1+k_3]\ne 0$. An explicit example is given by the choice
\begin{equation}
h(l)=(\epsilon^2+4a^2l^2)^{1/2}
\end{equation}
Once again, if we treat Rindler frame as a limit of a sequence of metrics, the scalar curvature is concentrated on the light cone and action is finite.

What we have done is to construct a class of metrics of the form $g_{ab}(\lambda, x^i)$ where $\lambda$ is a parameter, such that when $\lambda\to 0$
the metric reduces to flat spacetime in some curvilinear coordinates. But, in the same limit, the curvature scalar (as well as some of the components of the curvature tensor) becomes a distribution [like Dirac delta function]. Calculating the curvature by taking the derivatives of the limiting form of metric, gives a different result from calculating the curvature and then taking the limit.  

If $h(l)=2al$ in (\ref{classofmetrics}), then a coordinate transformation with $x=(2l/a)^{1/2}$ will convert (\ref{classofmetrics}) to
(\ref{startmetric}) with $f(x)=a^2x^2$. The  Euclidean continuation $(t\to t_E=it$) of this metric when $h(l)=2al, a>0$
has the correct signature only for $l>0$. Examining the nature of the conical singularity at the origin
we again conclude that $t_E$ is periodic and we can take $l>0$; in fact, this is again just the polar coordinates $(r,\theta)$
with $t_E=\theta/a; l=(1/2)ar^2$. Let us now consider a different metric with $h(l)=2a|l|$. For $l>0$, this is identical to
the metric with $h(l)=2al$; since the Euclidean extension cares only about $l>0$, the Euclidean extension will
again be the same! However, for  $h=2al, R=0$ while for $h=2a|l|,R=2a\delta(l)$. Thus the boundary condition at the origin
in Euclidean sector (or on the horizon in the Minkowski sector) is what distinguishes the two metrics with  $h=2al$
or $h=2a|l|$. Our earlier analysis shows that the metric in (\ref{startmetric}) with $f(x)=\epsilon^2+a^2x^2$ 
or the metric in (\ref{classofmetrics}) with $h(l)=(\epsilon^2+4a^2l^2)^{1/2}$,
provides a limiting procedure in which the $R\propto \delta_D(x^2)$ is picked
up.  The fact that length scales below Planck length cannot be operationally defined\cite{tplimit} makes this procedure particularly relevant.

These metrics play an important role in the in the complex path
approach\cite{complex} to quantum field theory near the horizon. The wave equation for a scalar field near the horizon can be reduced to a Schrodinger
equation with the potential $V(x)\propto -g_{00}|g^{11}|$.  Usually, if the horizon is at $x=0$, then
$g_{00}=-g^{11}\to (2ax)$ near the horizon, where $a$ is the surface gravity. This will lead to a singular
effective potential $V(x)\propto -x^{-2}$ and one needs to dip into complex-x plane to obtain finite 
results\cite{complex}. But for both classes of metrics considered above, 
 $g_{00}|g^{11}|=(\epsilon^2+b^2x^2)$ with some constant $b$, so that the effective potential reduces to
\begin{equation}
V(x)\propto -(\epsilon^2 + b^2 x^2)^{-1} 
\end{equation}
near the horizon with $\epsilon\to 0$. The poles of the potential are at $x=\pm i(\epsilon/b)$
and it is shown in the appendix of Ref.~\refcite{complex} that this regularisation of the $V(x)\propto -x^{-2}$ potential leads to the correct thermal behaviour for the horizon. Thus complex extension in either $t$ or $x$ possesses the same information
and the same regularisation works in both cases. [This issue is investigated fully in a separate publication].

\section{An electromagnetic analogy}

To provide some physical intuition into this bizarre situation, one may consider an analogy in the case of electrodynamics. In standard flat spacetime electrodynamics with vector potential $A_i(x^a)$,   a configuration of the form $A_i = \partial_i q(x^a)$ appears to be a pure gauge connection with
zero curvature. This result, of course, has well known caveats. If we take $x^a =(t, r, \theta, \phi)$ and
$q(x^a) = \phi$, then the vector potential $A_i = \partial_i(\phi)$ is \emph{not}
pure gauge and will correspond to a magnetic flux confined to a Aharanov-Bohm type solenoid
at the origin. The line integral of $A_idx^i$ around 
 the origin  will lead to a non-zero result, showing
 $\nabla \times {\bf A}$ is non zero
at the origin corresponding to $x^2 + y^2 =0$ in the Cartesian coordinates. In this case $q(x^a) =\phi =
\tan^{-1}(y/x)$. Instead, if we take $q(x^a) = \tanh^{-1}(t/x)$, then the same analysis goes through
leading to an electric field confined to the light cone $x^2 - t^2 =0$. This is seen most easily by noticing
that, in the Euclidean sector, there is no difference between $t_E-x$ plane and $x-y$ plane and 
   $\tan^{-1}(t_E/x)$ will go over to $\tanh^{-1}(t/x)$ as one proceeds from Euclidean electrodynamics
   to Lorentzian electrodynamics. Consider now an one-parameter class of  vector potentials
   of the form 
   \begin{equation}
   A^i (x^a; \epsilon)= \left( \frac{x}{x^2 - t^2 +\epsilon^2}, \, \frac{t}{x^2 - t^2 +\epsilon^2}, \, 0,\, 0\right)
   \label{vectorpot}
   \end{equation} 
   In the limit of $\epsilon\to 0$, this vector potential reduces to $A_i = \partial_i [\tanh^{-1}(t/x)]$.
   The electric field corresponding to the above vector potential has the form
   \begin{equation}
   {\bf E} = \left(\frac{-2\epsilon^2}{(x^2 - t^2 +\epsilon^2)^2},0,0\right)
   \label{electricfield}
   \end{equation}
   When $\epsilon\to 0$, this electric field becomes a distribution concentrated on the light cone:
   \begin{equation}
   E_x\to -2 \delta_D(x^2-t^2) 
   \label{elimit}
   \end{equation}
   Thus, one can construct electromagnetic field (curvature) concentrated on the light cone
   by a suitable limiting process.
   
   This situation is completely analogous to the Rindler frame example given above.
   The connection $(A_i)$ in (\ref{vectorpot}) is analogous to the gravitational connection
   \begin{equation}
   \Gamma^0_{0x} = \Gamma^0_{x0} = \frac{1}{2} \frac{f'}{f} = \frac{a^2 x}{\epsilon^2 + a^2 x^2}; \quad
   \Gamma^x_{00} = a^2 x
   \label{thegammas}
   \end{equation}
   for the metric in (\ref{startmetric}) with $f= \epsilon^2 + a^2 x^2$. When $\epsilon\to 0$, this connection can be 
   expressed as pure gauge (almost everywhere), just as $A_i$ in (\ref{vectorpot}) becomes $A_i = \partial_i q$. The 
   ${\bf E}$ in (\ref{electricfield})  is analogous to curvature in (\ref{reqn})  and 
   the limit in (\ref{elimit}) correspond to (\ref{rlimiteqn}).
   Just as the line integral 
   of $A_idx^i$ around the origin  indicates 
   the presence of a non trivial field configuration, one can construct the line integral
   over the gravitational connection to probe curvature. 
   In the case of asymptotically Rindler spacetimes, the Euclidean 
   metric will correspond to that of standard polar coordinates in the $T_E-X$ plane at large
   $x$. In this limit, the angular coordinate will correspond to $\theta =at_E$. The 
   line integral over a circle of large radius $x$ 
   \begin{equation}
   \oint_{x\to \infty} \Gamma^0_{0x}\sqrt{f}\, dt_E = \oint_{x\to \infty}\frac{1}{2} \frac{f'}{\sqrt{f}} dt_E
   \to 2\pi
   \end{equation}
   in the limit of $f\to a^2x^2$ indicating the existence of non zero curvature around the
   origin. 
   
    Just as the Aharanov-Bohm
    effect introduces non zero winding number in space, the concentrating of the scalar curvature
    in the origin of the Euclidean plane leads to a non zero winding number in the presence of 
    horizons and thermal effects
    (as originally suggested in Ref.~\refcite{chrisduff}). 
    Since the topological feature arises due to a circle of infinitesimal radius around the origin in the 
    Euclidean case, the analysis should work for any horizon which can be approximated by a Rindler
    metric near the horizon. 

\section{Aknowledgements}
    
    I thank Apoorva Patel for  several rounds of discussions,  S.V.Dhurandhar for  reference \refcite{stackgold}
and comments on distributions and J.V.Narlikar, Tulsi Dass, K.Subramanian
    for useful comments on the draft of the paper.


\section*{References}
\begin{thebibliography}{99}

 
 \bibitem{birreltp} Birrell N.D and Davies P.C.W, {\it Quantum fields in curved space}, (Cambridge University 
Press, Cambridge, 1982)
 
 
 
  
\bibitem {tprealms} Padmanabhan T.,  
       Mod.Phys.Letts. A {\bf 17}, 923 (2002). [gr-qc/0202078]. 
 
 
 \bibitem{chrisduff} Christensen, S.M and M.J. Duff, Nucl.Phys. {\bf B 146}, 11 (1978)

\bibitem{stackgold} 
 Stakgold, Ivar., {\it Green's functions and boundary value problems}, 
 John Wiley, New York (1979) p.110.
 
 
\bibitem{tplimit} 
T. Padmanabhan,
Class. Quant. Grav. {\bf 4} (1987) L107. 



\bibitem{complex} Srinivasan, K., T.Padmanabhan (1999) Phys.Rev. {\bf D 60}, 024007.


 

   
\end{thebibliography}
\end{document}